\pdfoutput=1
\documentclass[%
reprint,
superscriptaddress,
nofootinbib,
amsmath,amssymb,
aps,
]{revtex4-1}
\usepackage{graphicx}
\usepackage{bm}
\usepackage{hyperref}

\usepackage[utf8]{inputenc}
\usepackage{amsmath,amssymb}
\usepackage[T1]{fontenc} 
\usepackage{microtype} 
\usepackage[english]{babel} 

\usepackage{booktabs} 

\usepackage{graphicx}
\usepackage{amssymb}
\usepackage{braket,mleftright}
\usepackage{empheq}
\usepackage{subfigure}
\usepackage{stackrel}
\usepackage{soul}
\usepackage{blkarray}
\usepackage{multirow}
\usepackage{amsmath}
\usepackage{physics}
\usepackage{amsfonts}
\usepackage{bm}
\usepackage{bbold} 
\usepackage{color}
\newcommand{\beq}{\begin{equation}}
\newcommand{\eeq}{\end{equation}}
\newcommand{\bse}{\begin{subequations}}
	\newcommand{\ese}{\end{subequations}}
\newcommand{\bea}{\begin{eqnarray}}
\newcommand{\eea}{\end{eqnarray}}

\usepackage[utf8]{inputenc} 
\usepackage{amsmath}

\usepackage{enumitem} 
\setlist[itemize]{noitemsep} 

\usepackage{hyperref} 
\usepackage{braket}
\usepackage{verbatim} 

\begin{document}
	
	
	\title{Quantum Optimal Control: Landscape Structure and Topology}
	
	\author{Mart\'{i}n Larocca}
	\email{mail to: larocca@df.uba.ar}
	
	\affiliation{Departamento de F\'{i}sica “J. J. Giambiagi” and IFIBA, FCEyN, Universidad de Buenos Aires, 1428 Buenos Aires, Argentina
	}%
	\author{Pablo Poggi}
	\affiliation{Departamento de F\'{i}sica “J. J. Giambiagi” and IFIBA, FCEyN, Universidad de Buenos Aires, 1428 Buenos Aires, Argentina
	}%
	\affiliation{Center for Quantum Information and Control, University of New Mexico, MSC07-4220, Albuquerque, New Mexico 87131-0001, USA}
	\author{Diego Wisniacki}
	
	\affiliation{Departamento de F\'{i}sica “J. J. Giambiagi” and IFIBA, FCEyN, Universidad de Buenos Aires, 1428 Buenos Aires, Argentina
	}%

	\date{January, 2018}%
	
	\begin{abstract}
		The core problem in optimal control theory applied to quantum systems is to determine the temporal shape of an applied field in order to maximize the expectation of value of some physical observable. The functional which maps the control field into a given value of the observable defines a Quantum Control Landscape (QCL). Studying the topological and structural features of these landscapes is of critical importance for understanding the process of finding the optimal fields required to effectively control the system, specially when external constraints are placed on both the field $\epsilon(t)$ and the available control duration $T$. In this work we analyze the rich structure of the $QCL$ of the paradigmatic Landau-Zener two-level model, studying several features of the optimized solutions, such as their abundance, spatial distribution and fidelities. We also inspect the optimization trajectories in parameter space. We are able rationalize several geometrical and topological aspects of the QCL of this simple model and the effects produced by the constraints.  Our study opens the door for a deeper understanding of the QCL of general quantum systems. 
		
		
		
	\end{abstract}

	\maketitle
	
	
	\section{Introduction} \label{Section-Intro}
	
	The development of new technologies based on quantum information processing is living a sprouting era. Proposals for communication, computation and simulation protocols based on quantum mechanical effects \cite{bib:libro_qc,bib:gisin2007,bib:nori2014} are nowadays being transformed into reality thanks to the extraordinary capabilities of physical platforms such as ion traps, quantum dots and superconducting qubits \cite{bib:martinis2016,bib:yin2017,bib:lukin2017}. To take full advantage of this, scientists rely on their growing ability to control physical systems in the quantum regime by using properly tailored external fields. In this context, optimization methods originally put forward in the late 1980s have proven to give robust control strategies \cite{bib:rabitz1988,bib:tannor1993}. \\
		
	The typical problem to solve is to find the control field $\epsilon(t)$ which maximizes a certain objective functional $J[\epsilon]$, i.e. the probability of reaching a target state, for instance.	Extensive application and study of this quantum optimal control (QOC) techniques over the past decades gave evidence of the benign features of what is commonly called the quantum control landscape (QCL), that is, the functional dependence of the objective $J$ with the field $\epsilon$. In a seminal work \cite{bib:rabitz2004}, Rabitz \emph{et al.} showed that, under certain conditions, the QCL was devoid of sub-optimal local maxima, which explained the extraordinary success of local optimization procedures. This remarkable result about the topology of $QCL's$ has been intensively tested and studied over the past decades \cite{bib:shen2006,bib:hsieh2009,bib:moore2012}, and limitations are known to arise in certain cases. For instance, local maxima or traps are expected to appear when the control problem has constraints \cite{bib:pechen2011,bib:pechen2012,bib:dmitri2015,bib:dmitri2017}, for example due to the time-discretization of the fields or bandwidth and amplitude limitations imposed to them. Another interesting constrain is given by the evolution time of the system. Traps have been shown to exist in the vicinity of the minimal or quantum speed limit (QSL) time \cite{bib:sherson2016} \footnote{Here we use minimal time and QSL as synonymous of the shortest process duration with perfect control (fidelity equal to one). Other works consider QSL as a bound for the minimal evolution time between an initial and final states.}, and there have also been numerous reports of slowing down of optimization algorithms in that regime \cite{bib:gajdacz2015}. However, systematic analysis on how exactly these constraints affect the control landscape have been limited as to now, and a joint assessment of multiple types of constraints is currently lacking in the literature.\\
	
	In this work we present a systematic analysis of the effects of coarse grained temporal fields and restricted evolution time on the structural and topological features of QCLs. Note that for a single control field, the optimization space has a dimension of $N_{ts}$, which is the number of time slots we use to discretize our temporal variable. Typically, $N_{ts}$ may be of the order of $10^2$ or $10^3$, and so it is not trivial to asess the global properties of $J[\epsilon]$, whose representation is given by a hypersurface in a $N_{ts}+1$ dimensional space. We therefore propose a number of strategies to probe the QCL in order to obtain information about its features. By using random initial seeds we explore a certain region of the parameter space, and using standard local optimization techniques we arrive at optimized solutions. For those, we study i) the distances between them, which allows us to to probe the number and distribution of maxima in such region, ii) their fidelities, which give us information about the emergence of traps due to the constraints imposed to the problem, and iii) a structural parameter $R$ defined in \cite{bib:r1}, which measures how straight is the path between the initial seed to the optimized solution. This parameter allows us to observe structural properties of the landscape.\\
	
	We use as a testbed for our analysis a simple, yet paradigmatic model of a driven two-level quantum system which is described by the Landau-Zener Hamiltonian. For this model, a related study explicitly showed that the control landscape is indeed devoid of traps for $N_{ts}\rightarrow\infty$ \cite{bib:pechen2012}. Here we go beyond that result and characterize not only the topology of the landscape but also its geometrical structure as a function of both $N_{ts}$ and the evolution time $T$. Although local maxima (traps) disappear in the limit of continuous field, global maxima are shown to exhibit an interesting two clan structure in the vicinity of the Quantum Speed Limit ($QSL$). Moreover, this two families merge at the $QSL$, rendering only one (global) maximum for any landscape with $T<T_{min}$. Regarding the geometry of the landscape, we analyze the straightness of the trajectories transversed by a pure gradient algorithm towards the maxima, by comparing the actual path length and the euclidean distance between initial and optimized fields. These trajectories through control space are found to bend as $T$ approaches $T_{min}$, and a discontinuous jump is observed at $T=T_{min}$ where every path from hundreds of random seeds reach the only global maximum in a perfect straight line. That is, the landscape is found to be trivial at the quantum speed limit.\\
	
	This paper is organized as follows. In Section II we present the basics of optimal control theory and its application to the Landau Zener two-level model. In Section III we will present a ``toy model'' in which the control field is discretized into just $N_{ts}=2$ time steps. This will allow us to visualize the landscape directly, and thus will be helpful for designing strategies that allow us to probe its features in a more general setting. In Section IV we discuss such strategies and show results for general landscapes with $N_{ts}>2$. Finally, in Section V we present some concluding remarks. 
	
	\section{\label{SectionII} Quantum optimal control and the Landau-Zener Model}
	
	Consider the time evolution of an isolated driven quantum system described by the following Schr\"odinger equation
	\begin{equation}
	i\frac{d\hat{U_t}}{dt}=[\hat{H_0}+\epsilon(t)\hat{H_c}]\hat{U_t},
	\end{equation}
	
	\noindent where $\hat{U_t}$ is the unitary evolution operator of the system at time $t$, $H_0$ and $H_c$ are the drift and control Hamiltonians respectively and $\epsilon(t)$ symbolizes the control field. Note that we set $\hbar=1$ from here on. Optimal control theory assesses the problem of deriving the shape of $\epsilon(t)$ that maximizes the value of a cost functional $J[\epsilon]$. For example, a typical goal of control tasks is to take a given initial state $\ket{i}$ to a desired target state $\ket{f}$ in a (fixed) time $t=T$, that is to obtain $U_T\ket{i}=\ket{f}$ (up to some global phase). Finding the fields that perform the desired task with the best possible accuracy is identical to locating the global maxima of the QCL $J[\epsilon]$, which, in this particular case, would simply take the form $J[\epsilon]=|\bra{f}U_T\ket{i}|^2$. We can look for such maxima by proposing an initial seed $\epsilon_{(0)}(t)$ for the field, and to update it iteratively by using information about the gradient of $J[\epsilon]$. This is the idea behind most of the QOC methods which have been widely incorporated by quantum scientists in the last decades, such as Krotov \cite{bib:krotov1,bib:krotov2}, GRAPE \cite{bib:grape1} and others. An algorithm of this type would generate a path through the landscape which connects $\epsilon_{(0)}$ to some optimal field $\epsilon_{(K)}(t)$, where $K$ denotes the number of iterations. Note that while ideally we expect $J[\epsilon_{(K)}]=1$, in general the optimization will stop either when $J[\epsilon_{(K)}]=1-\delta$ or when the gradient of the cost functional vanishes, $\nabla J[\epsilon_{(K)}] \simeq 0$. \\
	
	Optimal control techniques have been applied to a variety of scenarios ranging diverse areas, and has been especially fruitful in quantum chemistry \cite{bib:shi1988,bib:krotov1} and quantum information related protocols \cite{bib:poulsen2010,bib:goerz2011,bib:gambetta2015}. Here we will focus on a simple but non-trivial model of a controlled quantum system. Let us consider a two-level system described by the Hamiltonian
	\begin{equation}
	\hat{H}(\epsilon(t)) = \frac{\Delta}{2}\hat{\sigma}_{x}+\epsilon(t)\hat{\sigma}_{z}
	\label{ec:hami_lz}
	\end{equation}
	
	\noindent with $\sigma_{x}$ and $\sigma_{z}$ are Pauli matrices. Parameter $\Delta$ is usually referred to as the energy gap since its measures the minimum separation between the eigenenergy branches of $H(\epsilon)$, which occurs at $\epsilon=0$. This model has been widely applied in quantum physics, as it describes non-adiabatic transitions \cite{bib:zener1932}, Landau-Zener-Stuckelberg interferometry \cite{bib:nori2010} and quantum phase transitions \cite{bib:zurek2005}. When the state is initially prepared as $\ket{\psi(t\rightarrow-\infty)}=\ket{0}$, choosing $\epsilon(t)=v\:t$ yields the famous Landau-Zener problem \cite{bib:zener1932,bib:landau1932}, for which an analytical formula can be drawn for the asymptotic probability of population transfer between $\ket{0}$ and $\ket{1}$ (the eigenstates of $\sigma_z$). Here, we are interested in achieving complete population transfer between those states, and we will often be also interested in minimizing the evolution time. Linear sweeping of the control parameter is not efficient in this context, since large evolution times (scaling as $\Delta^{-2}$ \cite{bib:poggi2013}) would be required in virtue of the adiabatic theorem. As a consequence, we will resort to optimization techniques in order to find an appropriate shape for $\epsilon(t)$ that maximizes
	\begin{equation}
	J[\epsilon]=|\bra{1}U_T[\epsilon]\ket{0}|^2
	\label{ec:funcional}
	\end{equation}
	
	\noindent for each fixed value of $T$. The issue of time-optimal control in this scenario was studied by Hegerfeldt \cite{bib:hegerfeldt2013}, who showed that there is a minimum control time which is given by
	\begin{equation}
	T=T_{min}=\frac{\pi}{\Delta}.
	\label{ec:qsl}
	\end{equation}
	
	This means that, for $T<T_{min}$ it is not possible to achieve full population transfer, i.e., $J[\epsilon]<1$ for all $\epsilon(t)$. Remarkably, the field shape which accomplishes the control task at $T=T_{min}$ is simply $\epsilon(t)=0$. 
	
	\section{\label{SectionIII} Looking at the landscape: toy model for the control}
	
	
	\begin{figure*}[t]
		\begin{center}
			\includegraphics[width=1.0\textwidth]{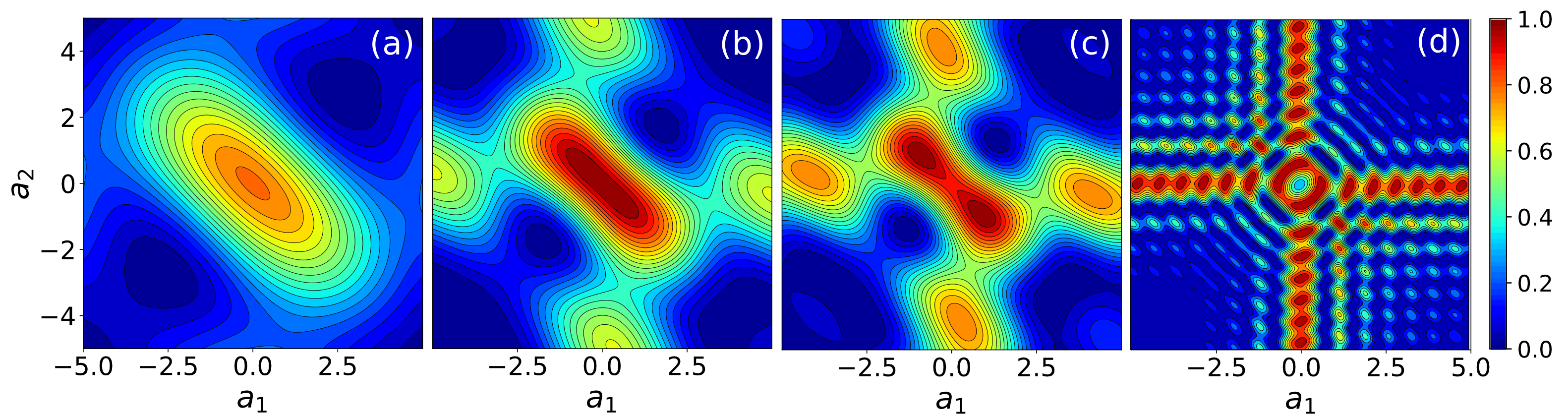}
				\caption{Control landscape for the Hamiltonian of Eqn. (\ref{ec:hami_lz}) with initial state $\ket{0}$ and target state $\ket{1}$. The case depicted here corresponds to scenario where $N_{ts}=2$ and so the control field is mapped to a two-dimensional vector $(a_1,a_2)$. Subplots correspond to different values of the total evolution time: (a) $T/T_{min}=0.7 $, (b)$T/T_{min}=1.0$, (c)$T/T_{min}=1.2$ and (d)$T/T_{min}=10 $. The energy gap is set to $\Delta=1$ in all cases. }
				\label{fig:contour}
		\end{center}
	\end{figure*}

    As already mentioned in the Introduction, it was shown in Ref. \cite{bib:pechen2012} that the control landscape $J[\epsilon]$ determined by the Landau-Zener Hamiltonian (\ref{ec:hami_lz}) has only global optima when we consider a continuous-in-time control field $\epsilon(t)$, that is, it is a trap-free model. Moreover, these global optima should correspond to $J[\epsilon]=1$ only for $T\geq T_{min}$. However, any practical realization of the optimization problem stated above implies performing a coarse graining of the temporal variable. As a result, the control function $\epsilon(t)$ will now be represented by a vector of control variables, namely
	\begin{equation}
	\epsilon(t)\rightarrow \left\{\epsilon_k\right\}\equiv \vec{\epsilon}
	\end{equation}
	\noindent with $k=1,2,\ldots,N_{ts}$. The functional dependence of $J$ is then mapped to an explicit dependence of the objective on the control parameters (and, also, on the evolution time)
	\begin{equation}
	J[\epsilon]\rightarrow J\left(\right\{\epsilon_k\left\},T\right)
	\end{equation}
	
	Naturally, for $N_{ts}>2$ we will not be able to visualize the landscape, and because of that in Section IV we will propose different strategies to obtain information about its features. In this Section we will study the simplest non-trivial scenario, where $N_{ts}=2$ in order to gain intuition about the properties of the landscape. For that, we propose that the field $\epsilon(t)$ is of the form  
	\begin{equation}
	\epsilon(t)=  \begin{cases} 
	a_1              & \mbox{if } t\leq T/2   \\
	a_2 & \mbox{if } t > T/2.
	\end{cases}
	\end{equation}
	
	It is then easy to evaluate the objective functional of eqn. (\ref{ec:funcional}),
	\begin{equation}
	J(a_1,a_2,T)=\lvert \Bra{1}e^{-iH(a_2)\frac{T}{2}}e^{-iH(a_1)\frac{T}{2}}\Ket{0}\rvert^2
	\label{ec:funcional_posta}
	\end{equation}
	
	In Fig. \ref{fig:contour}, we plot the control landscape of eqn. (\ref{ec:funcional_posta}), as a function of control parameters $(a_1,a_2)$ for different values of the total evolution time $T$, both below and above the quantum speed limit time $T_{min}$ [Eq. (\ref{ec:qsl})]. Fig. \ref{fig:contour}.(a) corresponds 
	to $T=0.7 T_{min}$. In this case, the landscape shows only a single global maximum  at the origin, with maximum fidelity of $J \simeq 0.65$ (sub-optimal). As $T \rightarrow T_{min}$ the landscape's topological structure remains unaltered, hosting a single global maximum (in the plotted region) with ever-growing fidelity. At $T=T_{min}$ the  global maximum reaches its optimal height $J=1$ [Fig.\ref{fig:contour} (b)].
	
	
	For $T>T_{min}$, a much more intricate structure in the topology of $J(a_1,a_2)$ is clearly observed in Fig.\ref{fig:contour} (c) and (d). The maximum at the origin splits into two symmetrical global maxima, which steadily separate from each other as the total evolution time is increased Fig.\ref{fig:contour} (c). In Fig. \ref{fig:contour} (d), we plot the control landscape at $T=10 T_{min}$. The landscape presents many extrema, as  was shown in Ref. \cite{bib:pechen2012}. We observe a shrinking of the characteristic scales as a function of control parameters $(a_1,a_2)$ for increasing $T$.\\
	
	
	It is important to point out that, in general, the minimum control time will be a function of $N_{ts}$. This is so because we defined $T_{min}$ as the minimum value of $T$ such that $J(\{\epsilon_k\},T)=1$. However, for the two-level problem considered here, analytical arguments have shown that $T_{min}=\pi/\Delta$ \cite{bib:hegerfeldt2013} and the corresponding optimal field is $\epsilon(t)=0$. Since that field is trivially achieved with any discretization (i.e., any value of $N_{ts}$), we don't have to worry here about having to consider a $N_{ts}$-dependent minimum control time.\\
	
	As we pointed out in the last Section, the strategy to obtain optimal fields using QOC is to propose an initial seed field and update it iteratively by using information about the gradient of $J[\epsilon]$. This process stops when 
	the solution converges to an extremum (with some error threshold). We can interpret
	this procedure as a path through the landscape. In Refs. \cite{bib:r1,bib:r2,bib:r3} a measure $R$ of the 
	straightness of the paths was proposed. $R$ is defined as the ratio between the length of the optimization trajectory, defined as
	\begin{equation}
	d_{PL}=\int_{0}^{s_{max}}\left[ \,\frac{1}{T} \int_{0}^{T} \left(\frac{\partial \epsilon(s,t)}{\partial s}\right)^2 dt\right] ^{1/2} ds,
	\label{dpl}
	\end{equation}
	and the Euclidean distance between the initial seed and optimal control field,
	
	\begin{equation}
	d_{EL}= \left[\frac{1}{T} \int_{0}^{T} \left[ \,\epsilon(s_{max},t)-\epsilon(0,t)] \, dt \right]\right]^{1/2} .
		\label{del}
	\end{equation}

	The new variable s parametrizes the optimization path, such that the initial seed is $\epsilon(s=0,t)$ and the optimized solution is given by $\epsilon(s=s_{max},t)$. We stress that the importance of inquiring into the non-topological details of the landscape is evident, since a trap-free landscape does not guarantee an easy optimization. Complex-structured landscapes may constrain optimization paths to inefficient twisted routes.\\
	
\begin{figure}
		\begin{center}
			\includegraphics[width=.5\textwidth]{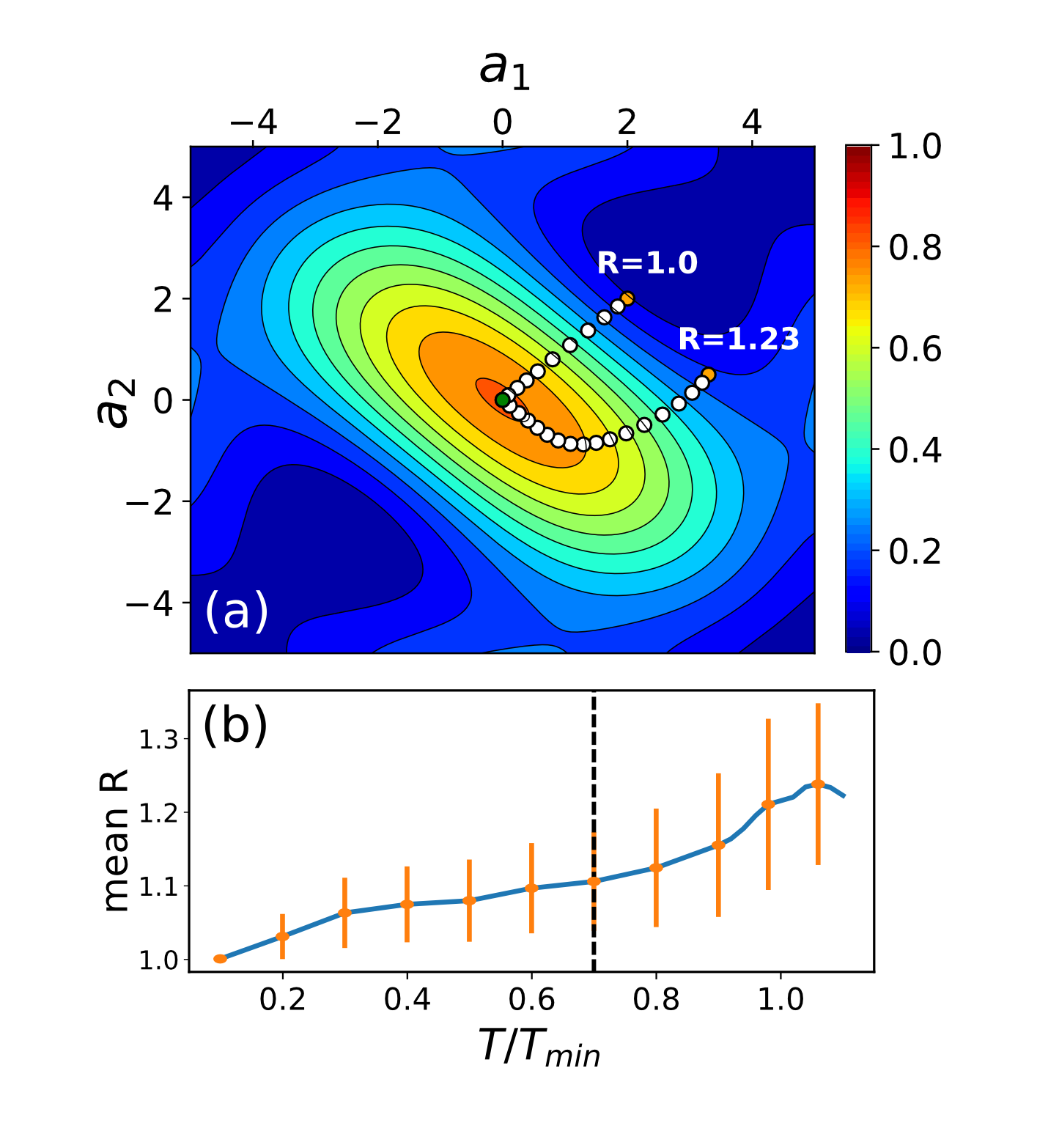}
				\caption{(Color online) (a) Two optimization trajectories in the landscape of Eq. (\ref{ec:funcional_posta}) for $T/T_{min}=0.7$. The corresponding initial seeds are shown as orange circles, and the final optimized parameters are shown as green circles. (b) Mean value of $R$ as a function of $T/T_{min}$ using 1000 initial seeds. Vertical dashed line indicate the value of $T/T_{min}$ in which the trajectories of panel (a) where computed.} 
				\label{Rnts2}
		\end{center}
	\end{figure}

	In Fig.  \ref{Rnts2} we show some results of the calculation of $R$ using a simple steepest ascent algorithm. Two representative trajectories through the landscape of Eq. (\ref{ec:funcional_posta}) for $T=0.8 T_{min}$ are plotted
	in Fig. \ref{Rnts2} (a). As we can see, one of them is completely straight, giving $R=1$
	whereas the other one, yielding $R=1.23$, is slightly arched \cite{bib:r1}. 
	In order to gain some insight of the structure of the landscape, a statistical analysis of $R$ was performed using $1000$ random initial seed fields in the region $-1<a_1,\:a_2<1$. Each initial field was optimized and the measure $R$ of its path in the landscape was computed. Mean value and standard deviation of the R distributions are shown in Fig. \ref{Rnts2} (b) for different values of $T$. Only those trajectories leading to a global maximum were considered. This figure 
	indicate that the lenght $R$ of the path towards the optimal and therefore its complexity is increased with $T/T_{min}$. This is consistent with the structure of the landscape that was plotted in Fig. \ref{fig:contour} (a-d).
	
	
	\section{\label{SectionIV} General control fields and multidimensional landscapes}
	
	As already mentioned, we cannot directly visualize the landscape for control space dimensions above $N_{ts}=2$. We can, however, obtain information about it by generating a large number of initial seeds and analyzing the resulting control trajectories statistically. This idea has been used in previous works on quantum optimal control \cite{bib:pechen2012,bib:sherson2016}. This approach has intrinsic limitations, since limited computational resources give rise to what is usually known as the exploration-exploitation trade-off in optimization theory \cite{bib:sherson2017}, by means of which  a \textit{detailed} characterization the \textit{whole} multidimensional landscape is out of reach. Here we will focus on probing the landscape in a region centered around $\epsilon(t)=0$. This is an obvious reference in this case, since the energy spectrum of the Hamiltonian in eqn. (\ref{ec:hami_lz}) is symmetric with respect to $\epsilon=0$. Also, using constant, feature-less fields as initial guesses is a common approach in optimal control problems. Finally, its worth pointing out that, as mentioned in Section II, this is the actual optimal field for $T=T_{min}$. Thus, the chosen landscape region is relevant for exploring.\\
	
	In the remainder of this section we propose different methods for probing the structure and topology of the control landscape. The common methodology is as follows. An initial guess for the control field $\epsilon^{(0)}(t)$ is generated as a vector of random numbers $\{\epsilon_k^{(0)}\}$, where $-A\leq\epsilon_k^{(0)}\leq A$ and $k=1,\ldots,N_{ts}$. The field is then optimized using  GRAPE algorithm, which is currently built-in in the QuTiP Python package \cite{bib:qutip,bib:qutip2}. The iterative optimization stops when the gradient of the functional satisfies an standard convergence criterion. The process is then repeated for a large number of random initial seeds (of the order of 1000) in order to draw sufficient statistics.
	
	\subsection{Distance between optimal fields}
	\label{SectionIVA}
	
	Our first focus of interest is on the topology of the landscape, namely, the number, distribution and nature of its extrema. When dealing with multidimensional control optimization, it is a well known fact that different initial guesses lead generally to different optimized fields, albeit usually yielding similar optimized fidelities. In order to explore this behavior, we study the distribution of the optimized control fields as follows. For each pair of optimal fields found, we calculate the distance between them simply as
	
	\begin{equation}
	D_{ij}=\frac{1}{T}\int_{0}^{T} \lvert\epsilon^{(i)}(t')-\epsilon^{(j)}(t')\:\rvert dt'
	\label{ec:distance}
	\end{equation}
	
\begin{figure}
	\begin{center}
		\includegraphics[width=\linewidth]{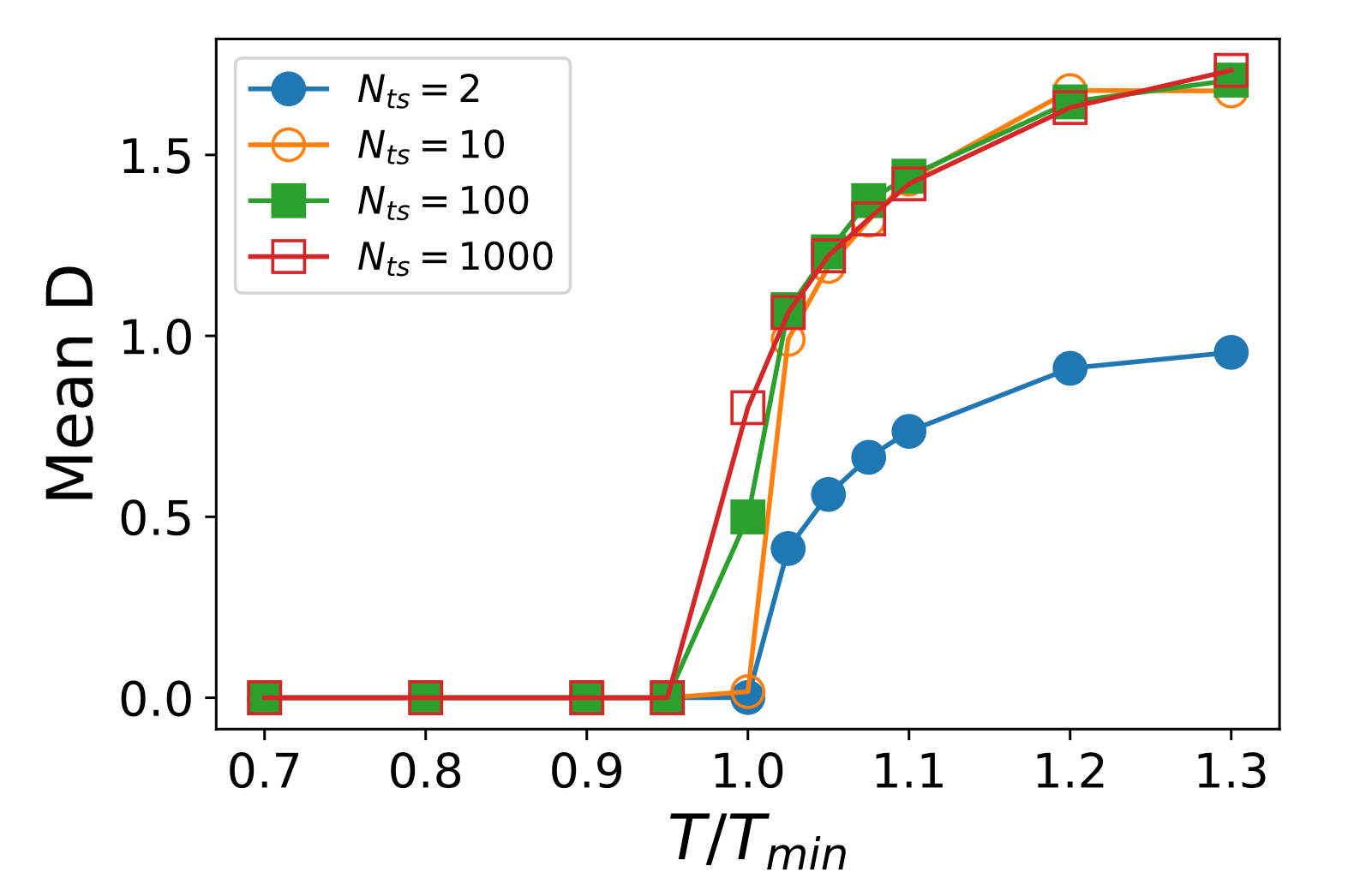}
		\caption{Mean value of the distance between optimal fields as defined in eqn. (\ref{ec:distance}), as a function of the total evolution time measured in units of $T_{min}=\pi/\Delta$. The results of a statistical analysis, involving one thousand random initial seeds in the $A=1$ region, are shown for different choices of $N_{ts}$, the number of time slots in which the field is discretized.}
		\label{fig:dis}
	\end{center}
	\end{figure}

	The mean value of the distance $\langle D\rangle $ between optimized fields is plotted in Fig. \ref{fig:dis} as a function of the evolution time $T$, and for different values of the number of time slots $N_{ts}$. There, we can see that $\langle D\rangle$ is close to zero when we intend to control the system below the quantum speed limit time, and rises steadily beyond that regime. This tells us that the control landscape for $T/T_{min}<1$ has essentially a single maximum to which all initial seeds converge. However, for longer control times, optimized solutions spread out in multiple global maxima. This is the same feature that was found for the simple case of $N_{ts}=2$, and we show here that it extends to $N_{ts}$ of the order of 1000.\\
	
	A remarkable conclusion from this analysis is that the landscape undergoes a sudden transformation at $T=T_{min}$. Note that, a priori, we expected the evolution time $T$ to impact the height of the extrema, by definition. Nevertheless, results shown here demonstrate a more profound topological change on the landscape when traversing the quantum speed limit, giving rise to multiple extrema which were absent for smaller control times. It would be interesting to explore whether this phenomenon takes place also in more complex quantum systems. We leave this issue for future work.\\
	
	We also point out that that the spreading of optimal solutions at the onset of controllability, i.e. for $T>T_{min}$, is consistent with the concept of \textit{superlandscape} introduced recently. This is due to the fact that small (time localized) perturbations of an optimum field can be easily compensated to make the perturbed field optimal as well, leading to many closely spaced locally optimal solutions \cite{bib:sherson2017}.\\
	
\begin{figure}
	\begin{center}
		\includegraphics[width=\linewidth]{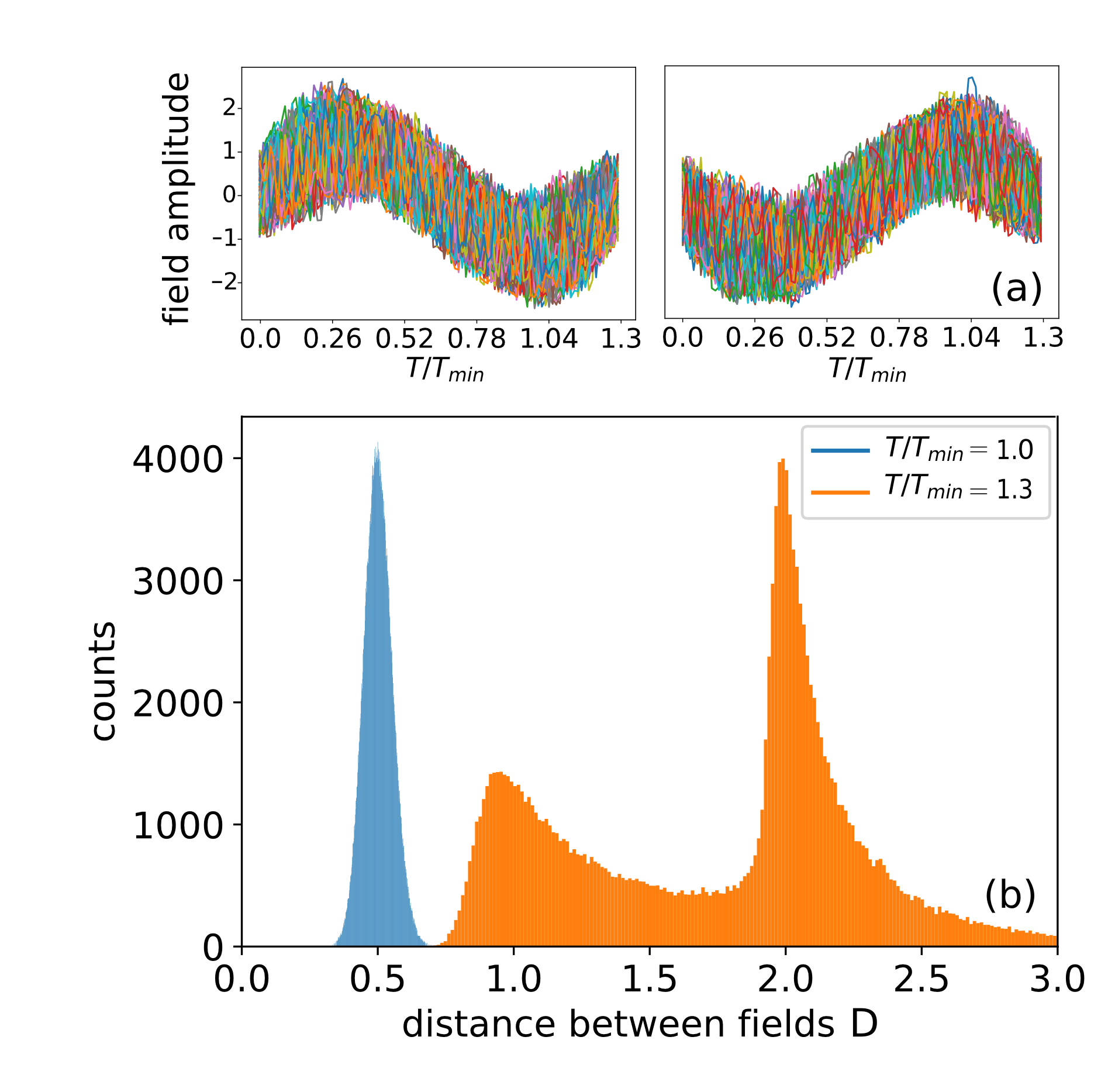}
		\caption{(a) Optimized control fields belonging to different regions (clusters) in control space $\vec{\epsilon}_A$ and $\vec{\epsilon}_B$ (see text for details). (b) Distribution of distances between control fields as calculated by eqn. (\ref{ec:distance}), for two different values of the evolution time $T$. The initial seed fields for optimization were generated as vectors of random numbers in the $A=1$ region. The number of time slots $N_{ts}$ was fixed to a hundred. Notice that histograms for $T/T_{min}<1$ have zero mean and vanishingly small variance and so are not shown in this plot.}
		\label{fig:histo}
	\end{center}
	\end{figure}

	In order to obtain a deeper insight about this result, we take a closer look at the actual distribution of distances found by this procedure. Results are shown in Fig. \ref{fig:histo} (b) for two different values of the evolution time $T$. For $T/T_{min}=1$, distances between optimized fields form a unimodal distribution, indicating a narrow spread of optimal solutions around some point in parameter space. As already seen in Fig. \ref{fig:dis}, for $T/T_{min}>1$ the mean distance shifts to larger values. Interestingly also, the distribution becomes bimodal in this case. This means that the optimized solutions now cluster around two points in control space $\vec{\epsilon}_A$ and $\vec{\epsilon}_B$; the leftmost peak in the distribution corresponds to distances between solutions in the same cluster, and the rightmost peak to solutions in different clusters. This clustering behavior can be understood as the emergence of the two global maxima in the superlandscape. Representative solutions of each type are plotted in Fig. \ref{fig:histo} (a), from which we observe that $\vec{\epsilon}_A\sim -\vec{\epsilon}_B$. This relation between optimal fields was already seen in the two-dimensional case $N_{ts}=2$, see Fig. \ref{fig:contour} (c). This shows a non trivial connection between the easily tractable low-dimensional control landscape and the complex multi-dimensional one.
	
	\subsection{Trapping probability}
	\label{SectionIVB}
	
\begin{figure}
	\begin{center}
		\includegraphics[width=.5\textwidth]{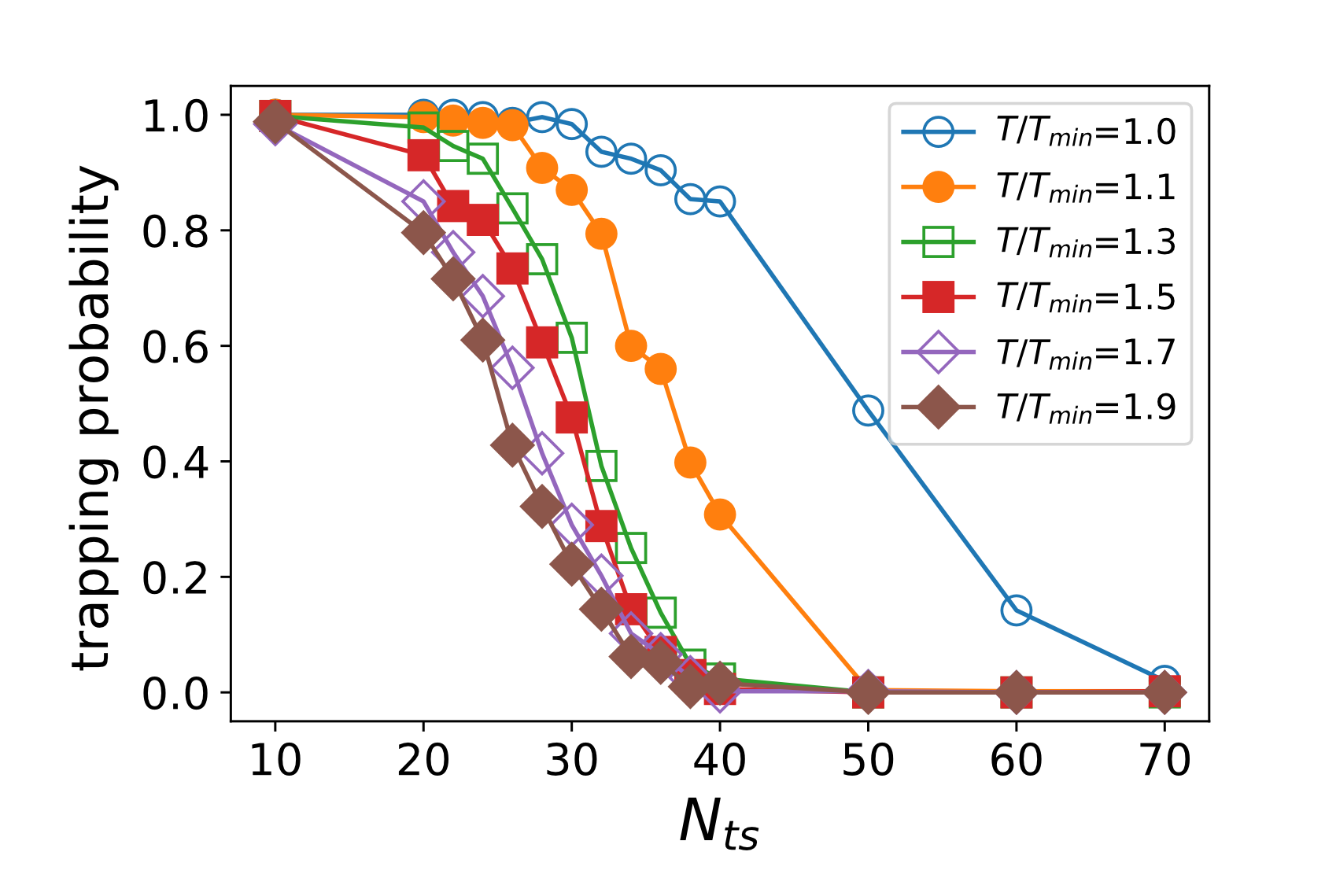}
		\caption{Trapping probability as a function of $N_{ts}$ (the number of parameters in the control field), for different values of the total evolution time $T$. A control trajectory is said to be trapped if it converges to a final fidelity below 0.99 (see text for more details). Random seeds were taken, now from the much bigger region of parameter space $A=50$}
		\label{fig:traps}
		\end{center}
	\end{figure}

	In this section we explore the emergence of traps, i.e. sub-optimal local maxima in the landscape, as a function of the constraints imposed on the control problem. In order to do this, we consider that a particular control trajectory has become trapped if it is unable to reach a final fidelity greater than $J=0.99$. Then, for fixed $T$ and $N_{ts}$ we define the trapping probability as the fraction of optimized solutions that became trapped.  In Fig. \ref{fig:traps} we plot this quantity as a function of the number of time slots $N_{ts}$, for different evolution times $T/T_{min}$. It is readily seen from the plot that the trapping probability goes to zero for large $N_{ts}$ in all cases. This had been shown in Ref. \cite{bib:pechen2012}, where the authors analytically proved that the in the limit $N_{ts}\rightarrow\infty$ the landscape is indeed devoid of traps.
	
	From Fig. \ref{fig:traps} it can be seen that for values of $T$ far from the quantum speed limit, the different curves approach each other, meaning there is no significant change in the abundance of traps with total time variation. However, when $T/T_{min}$ approaches 1, the behavior is markedly different, and the trapping probability decay slows down. This implies that on the onset of controllability the optimization becomes harder, as it is more likely for a random initial seed to converge to a trap. This interesting result highlights the role of the evolution time as an important constraint in control problems.
	
	We point out that many previous works have reported that optimal control near the quantum speed limit time tend to become \emph{slower}, in the sense that more iterations are needed in order to reach a satisfactory fidelity. In the landscape picture, this can be understood as maxima becoming more \emph{flat} (as can be seen for example in Fig. \ref{fig:contour} b). We stress that the result shown in this work is of a different nature, since here we observe the appearance of traps for sufficiently small times. Since we set a convergence criterion for the optimization based on the gradient of the functional, we can effectively distinguish actual traps from flat global optima.
	
	\subsection{Looking at the structure of the landscape using the measure $R$}
	
	\label{SectionIVC}
	
	As we mention previously, the optimization procedure can be seen as a
	travel on the landscape and the measure $R$ was proposed in Ref. \cite{bib:r1,bib:r2,bib:r3} to quantify the straightness of such optimization paths. $R$ is defined as the ratio between the length of the optimization trajectory $d_{PL}$ [Eq. \ref{dpl}] and of the 
	Euclidean distance $d_{EL}$ [Eq. \ref{del}] between the initial seed and optimal control field. Our goal now is to see the behavior of $R$
	when $N_{ts} > 2$ so neither the landscape nor optimization paths can be directly visualized. 
	Instead, we probe the structural (non-topological) features of the landscape by analyzing the behaviour of the $R$ distributions.
	
\begin{figure}[!htb]
		\begin{center}
			\includegraphics[width=.5\textwidth]{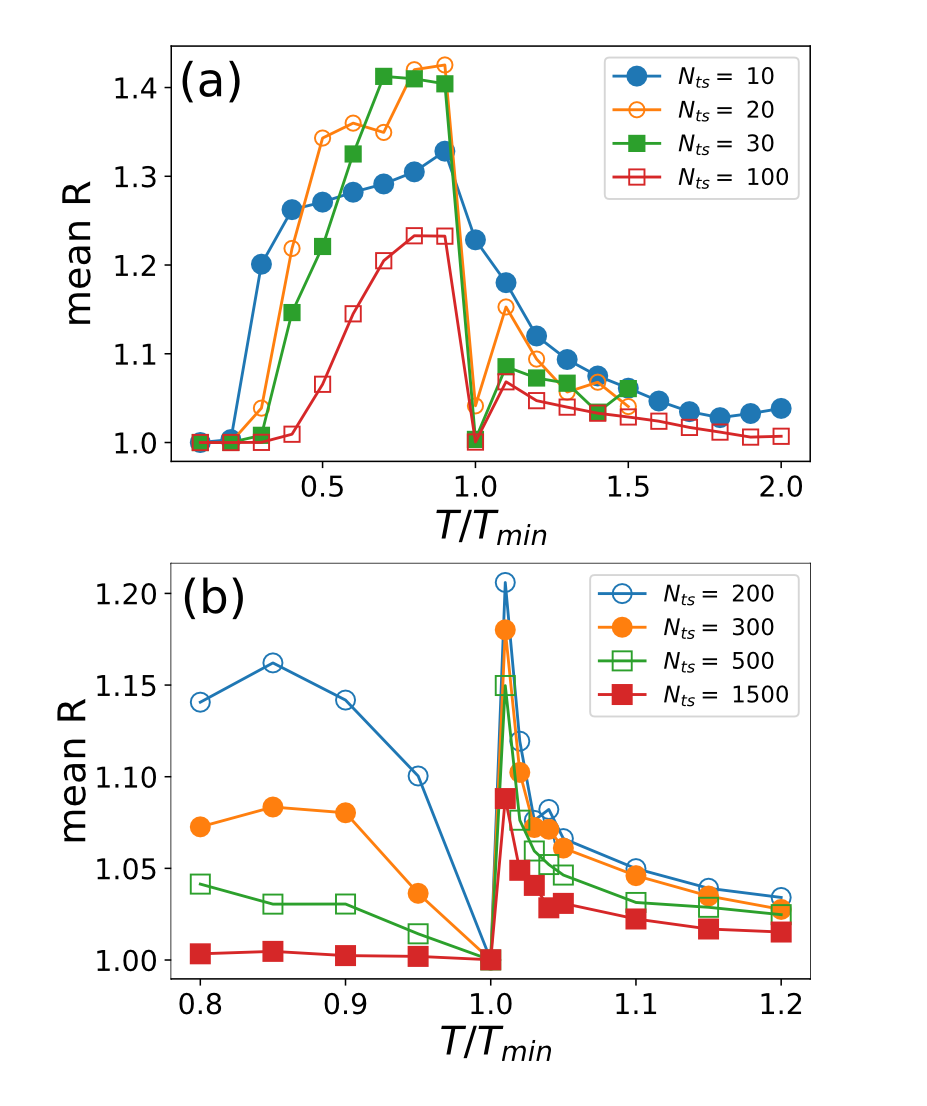}
			\begin{footnotesize}
				\caption{Mean value of $R$ as a function of $T/T_{min}$ for several number of times slots $N_{ts}$ of the control field. We have used 1000 initial seeds. (a) $N_{ts} = 10, 20, 30 $ and $100$. (b) $N_{ts} = 200, 300, 500 $ and $1000$.}
				\label{Rdip}
			\end{footnotesize}
		\end{center}
	\end{figure}

	In Fig. \ref{Rdip} we show the mean value of $R$ computed for $1000$ initial random seeds for several values of time slots $N_{ts}$ of the field. Several important features can be remarked. For small $T/T_{min} \rightarrow 0$ in  Fig. \ref{Rdip} (a) we can see that the mean $R$ approach to 1. This means that almost all trajectories  of the optimization process are straight lines showing that
	that the landscape  has a simple topological structure as was shown for $N_{ts} = 2$ in Fig.
	\ref{fig:contour} (a). Remarkably, the same is observed for  $T = T_{min}$
	and for $N_{ts} >10 $ (see the the pronounced deep trench in $T/T_{min}=1$). That is, the landscape has also a simple structure in $T_{min}$. For greater optimal evolution time $T$ the mean $R$ shows a sharp jump that indicates the birth of the two optimal solutions  shown in Fig. \ref{fig:histo}. We note that as $T$ is increased after the $T_{min}$, the mean value of $R$
	decays smoothly. This fact shows that the structure of the landscape is simpler when the constraint on the evolution time $T$ is relaxed, as expected \cite{bib:rabitz2004,bib:r1}. 
	
	In Fig. \ref{Rdip} (b) we show the mean value of $R$ for greater values of time slots $N_{ts} = 200, 300, 500 $ and $1000$.
	We can clearly see that the greater is $N_{ts}$, smaller $R$ is. This means that the optimization process becomes simpler as we increase the number of times slots of the field.
	


	\renewcommand{\figurename}{Figure} 


	\section{\label{Section-conclusion}Concluding remarks}
	
	
	The quantum control landscape is a functional that connects 
	a control field to a given value of an observable, and its structure determines the optimization process. In fact, such an optimization can be seen as a trajectory on the landscape, which is a multidimensional mathematical object that in general can not be 
	displayed.
	
	In this paper we study the quantum control landscape of a paradigmatic system: the two dimensional Landau-Zener Hamiltonian. We have made a systematic study of the influence of two important constraints: the discretization  and the time extension of the control field. When the number of time steps of the control field is two, the landscape can be directly plotted and several interesting features are easily visualized. In particular, near the minimum time at which the system can be controlled. For control fields with more than two time slots, we use indirect methods to unravel the structure of the landscape.
	We consider the distances between optimal fields, which allows us
	to map the topological structure of the landscape (for e.g. number and distribution of maxima). Regarding the fidelities of those optimal fields, we can test the emergence of traps due to the constraints imposed to the problem. Note that without constraints the system has been analytically demonstrated to be trap free.
	We also compute the $R$ metric defined in \cite{bib:r1}, which is a measure of how straight is the path taken by the optimizer along the landscape, connecting an initial seed with the corresponding optimized solution. This measure gives us information about the structural properties of the landscape. We remark that the landscape's topological behaviour around the minimum control time, which we directly observed for two time steps, is apparently maintained in higher dimensional landscapes (see Fig. \ref{fig:dis}).
    
    Considering that the quantum control landscape contains the relevant information
    for coherent control, and that we have been able to unravel its structure for a simple system,
    our work opens the door to the understanding of this important functional in more complex systems, specially its behaviour near the minimum control time. This is fundamental because it paves the way for the systematic generation of high-speed protocols that can effectively control real quantum systems facing decoherence.

	
	
	
	\begin{acknowledgements}
		P.M.P. acknowledges interesting discussions with Jens-Jakob Sorensen during the GRC Conference of Quantum Control in South Hadley. We would also like to thank Alexander Pitchfork of the QuTiP team and Gabriele De Chiara for the fruitful insights.
		This work was partially supported by CONICET Grant No. PIP 112201 501004 93CO, UBACyT Grant No. 20020130100406BA and National Science
		Foundation Grant No. PHY-1630114.

	\end{acknowledgements}

\bibliography{QCL.bib}

\end{document}